%% file: main.tex
\documentclass[11pt]{article}
\pdfoutput=1 

\usepackage[utf8]{inputenc}
\usepackage[T1]{fontenc}
\usepackage{lmodern}
\usepackage[margin=1in]{geometry}
\usepackage{microtype}
\usepackage{booktabs}
\usepackage{graphicx}
\usepackage{amsmath}
\usepackage{amssymb}
\usepackage{xcolor}
\usepackage{caption}
\usepackage{enumitem}
\usepackage{tikz}
\usetikzlibrary{arrows.meta,positioning,fit,backgrounds,calc}
\usepackage[hidelinks]{hyperref}
\usepackage{url}
\usepackage{xurl} 

\IfFileExists{figures/numbers.tex}{\input{figures/numbers.tex}}{%
  \newcommand{\yzph}{\textbf{??}}%
  \newcommand{\yzWERtiny}{\yzph}\newcommand{\yzWERsmall}{\yzph}%
  \newcommand{\yzRTFsmall}{\yzph}%
  \newcommand{\yzSpeedTiny}{\yzph}\newcommand{\yzSpeedSmall}{\yzph}%
  \newcommand{\yzOverhead}{\yzph}%
  \newcommand{\yzCmdAcc}{\yzph}\newcommand{\yzCmdAsCmd}{\yzph}\newcommand{\yzCmdFPR}{\yzph}%
  \newcommand{\yzCmdMs}{\yzph}\newcommand{\yzCmdN}{\yzph}\newcommand{\yzDictN}{\yzph}%
  \newcommand{\yzVADspeech}{\yzph}\newcommand{\yzVADsilence}{\yzph}\newcommand{\yzVADmargin}{\yzph}%
  \newcommand{\yzDysFP}{\yzph}\newcommand{\yzDysRecall}{\yzph}\newcommand{\yzDysNclean}{\yzph}%
  \newcommand{\yzDysNdys}{\yzph}\newcommand{\yzTestFiles}{\yzph}\newcommand{\yzTestFuncs}{\yzph}%
  \newcommand{\yzADRs}{\yzph}\newcommand{\yzSLOC}{\yzph}\newcommand{\yzSrcFiles}{\yzph}%
  \newcommand{\yzCPUModel}{\yzph}\newcommand{\yzNumCPUs}{\yzph}\newcommand{\yzRAMGB}{\yzph}%
  \newcommand{\yzOSName}{\yzph}\newcommand{\yzPyVer}{\yzph}\newcommand{\yzFWVer}{\yzph}%
  \newcommand{\yzWERn}{\yzph}\newcommand{\yzWERaudioMin}{\yzph}%
}

\newcommand{\yazses}{\textsc{YazSes}}
\newcommand{\code}[1]{\texttt{#1}}

\title{\yazses{}: An Offline, Privacy-First, Cross-Platform\\ Hold-to-Talk Voice-Dictation System}
\author{Mohsen Seyedkazemi Ardebili\\
  \small NovaFabric\\
  \small \texttt{mohsen.seyedkazemi@gmail.com}}
\date{July 2026}

\begin{document}
\maketitle

\begin{abstract}
Cloud voice-dictation services deliver strong accuracy but require streaming a
user's speech to a remote provider, an unacceptable trade-off in privacy-sensitive
professions and offline or air-gapped settings; the leading on-device alternatives
are either platform-locked or aimed at expert scripting rather than plug-and-play
dictation. We present \yazses{}, an open-source (Apache-2.0) hold-to-talk voice
dictation daemon that runs entirely on-device, with a single codebase targeting
Linux, macOS, and Windows through a protocol-based platform abstraction.
\yazses{} transcribes speech locally with \code{faster-whisper} (CPU, int8) and
injects the result into the focused application; a fast regex command grammar,
backed by an optional small-language-model router, maps utterances to editor and
terminal actions. Nothing leaves the machine: recording is push-to-talk rather than
always-listening, there is no telemetry, and an opt-in personalization loop keeps its
corpus encrypted on-device and proposes configuration changes instead of shipping
data out. We describe the system
architecture---a staged pipeline behind a protocol-based platform abstraction with a
JSON-RPC control plane---and its privacy and threat model. We evaluate the shipping
Python implementation on a single commodity Linux laptop (\yzCPUModel{}); the macOS
and Windows backends are implemented and unit-tested but not end-to-end evaluated
here. On
\yzWERn{} LibriSpeech \emph{test-clean} utterances, word error rate ranges from
\yzWERtiny\% (\code{tiny.en}) to \yzWERsmall\% (\code{small.en}) at
$\text{RTF}\approx\yzRTFsmall$ for \code{small.en} (real-time-factor; decoding
faster than real time on CPU with no GPU). The command grammar reaches
\yzCmdAcc\% action accuracy with a \yzCmdFPR\% false-positive rate on plain
dictation at \yzCmdMs~ms per call, and the non-decode pipeline adds
\yzOverhead~ms of overhead. The system and the benchmark harness behind every number
in this paper are public at \url{https://github.com/MSKazemi/yazses}.
\end{abstract}

\section{Introduction}

Voice dictation has become a mainstream input modality, but the most accurate
consumer options are cloud services: the user's audio is streamed to a remote
provider for transcription. For many users this is disqualifying. Lawyers,
clinicians, journalists protecting sources, and researchers under data-handling
agreements often cannot send speech---which may contain privileged or regulated
content---to a third party. Users on air-gapped or intermittently connected
machines cannot rely on a network round-trip at all. And the recent trend toward
always-capturing ``memory'' features on the desktop, exemplified by Microsoft
Recall~\cite{msrecall}, has sharpened public concern that ambient capture of
on-screen and spoken activity is a security liability rather than a
convenience.

The offline alternatives each leave a gap. Dragon~\cite{dragon} is a mature,
paid product effectively limited to Windows. Windows Voice
Access~\cite{windowsvoiceaccess} is capable but Windows-only and closed. Talon
Voice~\cite{talon} is cross-platform and offline but is built for deep,
scriptable voice control with a learning curve, rather than plug-and-play
dictation. On Linux in particular there is no well-supported, turnkey, offline
dictation-\emph{and}-commands tool.

We present \yazses{}, an open-source system that fills this gap. Its design
commitments are: (i)~\textbf{on-device by construction}---all speech recognition
runs locally on the CPU via \code{faster-whisper}~\cite{fasterwhisper} built on
CTranslate2~\cite{ctranslate2}, with no GPU, API key, or account; (ii)~\textbf{push-to-talk,
not always-listening}---audio is captured only while a key is held, so there is no
ambient recording; (iii)~\textbf{cross-platform from one codebase}---Linux, macOS,
and Windows behind a single protocol-based abstraction; and (iv)~\textbf{privacy as
an invariant, not a setting}---zero telemetry, and an opt-in personalization loop
whose data is encrypted, stays local, and yields configuration proposals rather
than exported data or uploaded model weights.

This paper makes the following contributions:
\begin{itemize}[leftmargin=1.2em,itemsep=2pt]
  \item A description of the \yazses{} \textbf{system architecture}: a staged
  dictation pipeline, a protocol-based cross-platform abstraction, a JSON-RPC
  control plane, and platform-appropriate text-injection backends
  (Section~\ref{sec:design}).
  \item A \textbf{privacy and threat-model analysis} of an offline dictation
  daemon, including the invariants that make ``nothing leaves the machine''
  enforceable and a STRIDE-style enumeration of residual risks
  (Section~\ref{sec:privacy}).
  \item A design for \textbf{privacy-preserving on-device personalization}: an
  encrypted local learning corpus and a held-out-validated tuner that proposes
  configuration changes and never trains or uploads weights
  (Section~\ref{sec:personalization}).
  \item An \textbf{empirical evaluation} of the shipping implementation on
  commodity hardware---word error rate and real-time factor across model sizes,
  command-grammar precision, pipeline latency, and memory footprint---together
  with a released, reproducible benchmark harness (Section~\ref{sec:eval}).
\end{itemize}

\section{Related Work}
\label{sec:related}

\paragraph{On-device speech recognition.}
\yazses{} builds on Whisper~\cite{radford2023whisper}, an encoder--decoder ASR
model trained with large-scale weak supervision that transfers well across
domains. We use the \code{faster-whisper}~\cite{fasterwhisper} reimplementation
over CTranslate2~\cite{ctranslate2}, which runs int8-quantized Whisper on the CPU
at practical speed. Streaming-oriented models such as Moonshine~\cite{jeffries2024moonshine}
target low-latency live transcription and voice commands; an earlier exploratory
branch of \yazses{} investigated a dual-stack streaming design, but the shipping
system deliberately uses a single push-to-talk decode for simplicity and
robustness (Section~\ref{sec:limitations}).

\paragraph{Voice dictation and control systems.}
Table~\ref{tab:compare} situates \yazses{} among widely used dictation tools.
Dragon~\cite{dragon} and Windows Voice Access~\cite{windowsvoiceaccess} are
accurate but platform-locked and closed-source. Talon Voice~\cite{talon} is the
closest offline, cross-platform peer, but its strength---a rich scripting model
for power users---is a different design point from \yazses{}' plug-and-play
dictation with a fixed practical command grammar. Wispr Flow~\cite{wisprflow} is a
polished cloud service and thus off-device by design. \yazses{} is, to our
knowledge, the only tool in this set that is simultaneously open-source,
cross-platform (including first-class Linux), and fully offline.

\paragraph{Privacy of desktop capture.}
The privacy hazards of always-capturing desktop features~\cite{msrecall} motivate
\yazses{}' push-to-talk, zero-telemetry stance. Where those systems continuously
snapshot activity, \yazses{} records only while a key is held and retains nothing
by default.

\paragraph{Supporting components.}
Voice-activity detection gates silent buffers; production-grade neural VADs such as
Silero~\cite{sileovad} exist, and \yazses{} can optionally use one, but its default
gate is a calibrated RMS threshold (Section~\ref{sec:eval}). Speaker embeddings
(ECAPA-TDNN~\cite{desplanques2020ecapa}) underpin \yazses{}' optional, off-by-default
speaker features. Optional offline text cleanup and command routing use small
language models via grammar-constrained decoding~\cite{llamacpp}. We evaluate WER
using the standard Whisper text normalizer~\cite{whispernormalizer} and
JiWER~\cite{jiwer} on LibriSpeech~\cite{panayotov2015librispeech}.

\begin{table}[t]
\centering
\small
\caption{\yazses{} among common dictation tools. ``Offline'' = transcription runs
on-device with no network dependency. Claims reflect publicly documented behaviour.
Talon's core is free; some beta features are paid. For \yazses{},
(\checkmark)~=~implemented and unit-tested in CI but \emph{not end-to-end validated
in this work}; only Linux is (Section~\ref{sec:limitations}).}
\label{tab:compare}
\begin{tabular}{lccccc}
\toprule
 & \yazses{} & Dragon & Talon & Win.\ VA & Wispr \\
\midrule
Offline / on-device & \checkmark & \checkmark & \checkmark & \checkmark & --- \\
Linux               & \checkmark & ---        & \checkmark & ---        & --- \\
macOS               & (\checkmark) & ---      & \checkmark & ---        & \checkmark \\
Windows             & (\checkmark) & \checkmark & \checkmark & \checkmark & \checkmark \\
Voice commands      & \checkmark & \checkmark & \checkmark & \checkmark & limited \\
Open source         & \checkmark & ---        & ---        & ---        & --- \\
No subscription     & \checkmark & ---        & \checkmark & \checkmark & --- \\
\bottomrule
\end{tabular}
\end{table}

\section{System Design}
\label{sec:design}

\yazses{} is a long-lived background \emph{daemon} that owns a single hold-to-talk
hotkey and runs the dictation pipeline; a thin command-line tool and optional tray
form a control plane that talks to the daemon over local IPC. The model is
deliberately simple for the user: hold the key, speak, release, and the text
appears in the focused window shortly after release---the delay is dominated by the
decode and so scales with utterance length and model size (Section~\ref{sec:eval}).

\subsection{The dictation pipeline}

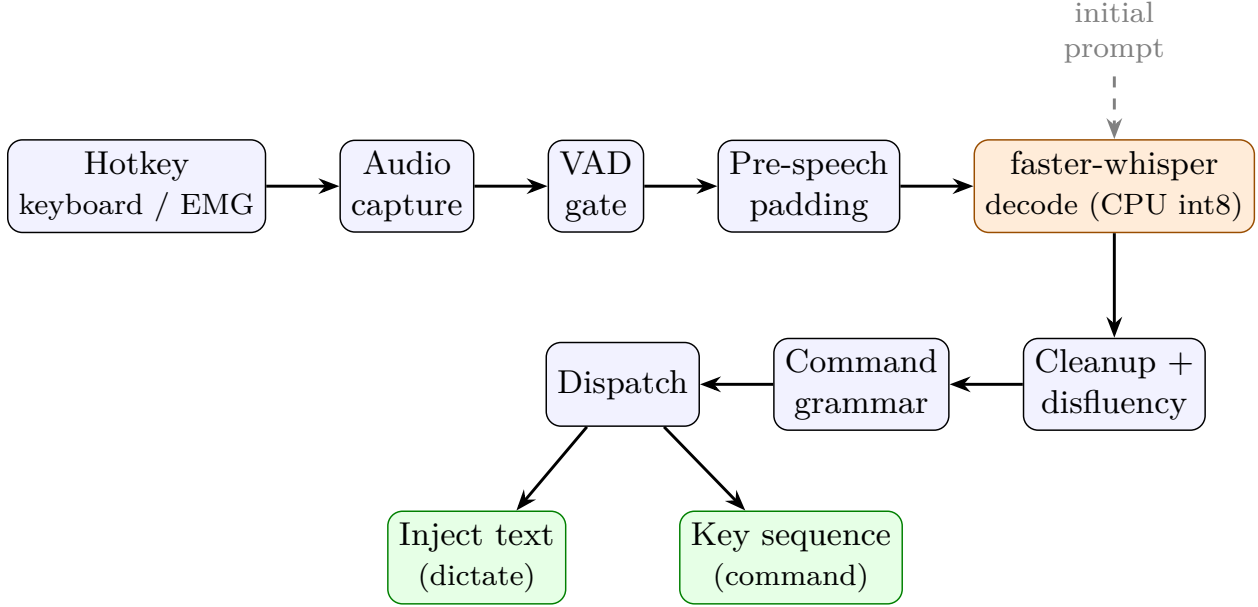
\begin{figure*}[t]
\centering
\resizebox{\textwidth}{!}{%
\begin{tikzpicture}[
  node distance=6mm and 7mm,
  box/.style={draw, rounded corners, align=center, minimum height=8mm,
    inner sep=3pt, font=\footnotesize, fill=blue!5},
  heavy/.style={box, fill=orange!15, draw=orange!60!black},
  term/.style={box, fill=green!10, draw=green!50!black},
  arr/.style={-{Stealth[length=2mm]}, thick},
]
\node[box] (hotkey) {Hotkey\\{\scriptsize keyboard / EMG}};
\node[box, right=of hotkey] (audio) {Audio\\capture};
\node[box, right=of audio] (vad) {VAD\\gate};
\node[box, right=of vad] (pad) {Pre-speech\\padding};
\node[heavy, right=of pad] (stt) {faster-whisper\\{\scriptsize decode (CPU int8)}};
\node[box, below=10mm of stt] (clean) {Cleanup +\\disfluency};
\node[box, left=of clean] (grammar) {Command\\grammar};
\node[box, left=of grammar] (disp) {Dispatch};
\node[term, below left=8mm and -2mm of disp] (inject) {Inject text\\{\scriptsize (dictate)}};
\node[term, below right=8mm and -2mm of disp] (keys) {Key sequence\\{\scriptsize (command)}};

\draw[arr] (hotkey) -- (audio);
\draw[arr] (audio) -- (vad);
\draw[arr] (vad) -- (pad);
\draw[arr] (pad) -- (stt);
\draw[arr] (stt) -- (clean);
\draw[arr] (clean) -- (grammar);
\draw[arr] (grammar) -- (disp);
\draw[arr] (disp) -- (inject);
\draw[arr] (disp) -- (keys);
\node[above=6mm of stt, font=\scriptsize, text=gray, align=center] (ip) {initial\\prompt};
\draw[arr, gray, dashed] (ip) -- (stt.north);
\end{tikzpicture}%
}
\caption{The \yazses{} dictation pipeline. Only the \code{faster-whisper} decode
(orange) is compute-heavy; every other stage is a pure transform costing well under
a millisecond (Table~\ref{tab:pipeline}). An \emph{initial prompt} (app name,
personal vocabulary, optional editor context) biases the decode.}
\label{fig:pipeline}
\end{figure*}

Each hold--speak--release cycle drives audio through the fixed sequence of stages in
Figure~\ref{fig:pipeline}. All stages except the Whisper decode are pure, fast
transforms.

The stages are: (1)~a \textbf{hotkey backend} (a keyboard hook, or an EMG
muscle-sensor over USB serial for hands-free/accessibility use); (2)~\textbf{audio
capture} to an in-memory buffer; (3)~a \textbf{VAD gate} that discards near-silent
buffers using a calibrated RMS threshold; (4)~\textbf{pre-speech padding} that
prepends a short lead-in so the first word is not clipped; (5)~the
\textbf{\code{faster-whisper} decode}, biased by an \code{initial\_prompt}
assembled from the application name, the user's personal vocabulary, and (optionally)
editor context; (6)~\textbf{text cleanup} that strips Whisper artefacts and a
three-pass \textbf{disfluency filter}; (7)~\textbf{command classification} by a
Tier-1 regex grammar, with an optional Tier-2 small-language-model router invoked
only when Tier-1 is unsure; and (8)~\textbf{dispatch}, which either injects the text
(dictation) or sends a key sequence (command).

\subsection{Cross-platform abstraction}

A single codebase serves three operating systems through protocol interfaces
(\code{HotkeyBackend}, \code{InjectorBackend}, \code{LifecycleBackend}, IPC,
permissions, tray). A factory selects concrete backends by operating system
(Table~\ref{tab:platform}). Adding a platform means implementing the protocols and
registering one \code{sys.platform} value; the daemon and CLI are unchanged. The EMG
hotkey is platform-independent and is registered whenever an EMG device is
configured.

Linux is the reference platform: it is the one used daily by the author and the only
one evaluated in this paper (Section~\ref{sec:eval}). The macOS and Windows backends
are implemented and exercised by the unit-test suite in a three-OS continuous-integration
matrix (last fully green at the v1.4.1 release), and installers for both are published
as unsigned developer previews---but we have not validated end-to-end dictation on
macOS or Windows in this work, and we flag this explicitly as a limitation
(Section~\ref{sec:limitations}).

\begin{table}[t]
\centering
\small
\caption{Protocol-based platform abstraction: one interface, three implementations.
This table describes the code; only the Linux column is end-to-end validated in this
paper (Section~\ref{sec:limitations}).}
\label{tab:platform}
\begin{tabular}{llll}
\toprule
Interface & Linux & macOS & Windows \\
\midrule
Hotkey     & evdev            & Quartz tap & keyboard hook \\
Injector   & ydotool/xdotool  & Quartz     & SendInput \\
Lifecycle  & systemd          & launchd    & SCM \\
IPC        & Unix socket      & Unix socket& named pipe \\
Tray       & ---              & rumps      & pystray \\
\bottomrule
\end{tabular}
\end{table}

\subsection{Control plane and text injection}

The CLI and tray never touch the pipeline; they drive the daemon over
\textbf{JSON-RPC 2.0} on a Unix domain socket (Linux/macOS) or named pipe (Windows),
carrying status queries, lifecycle commands, and correction signals. The daemon is a
state machine (\code{LOADING}$\rightarrow$\code{IDLE}$\leftrightarrow$\code{RECORDING}
$\rightarrow$\code{TRANSCRIBING}$\rightarrow$\code{INJECTING}), with additional
states for remote setup and enrollment.

Text injection is itself an abstraction with runtime backend selection. On Linux,
Wayland compositors block synthetic-input tools such as \code{wtype} for security;
\yazses{} therefore defaults to \code{ydotool}~\cite{ydotool}, which injects through
the kernel \code{uinput} device and works on Wayland and in terminals, falling back
to \code{xdotool} on X11 or a clipboard-paste path where appropriate. A
\textbf{remote path} forwards \emph{only the final text} to an agent on an SSH host,
so voice can drive a remote session while the audio never leaves the local machine.

\section{Privacy and Threat Model}
\label{sec:privacy}

\yazses{} treats privacy as an invariant enforced by construction, not a
configurable preference. The core invariants are: recording is push-to-talk, so no
audio is captured while the key is up; there is no telemetry or analytics of any
kind; no audio or text is transmitted off-device on the default path; the optional
remote path transmits final text only; and the optional learning corpus is encrypted
at rest with a machine-bound key and never leaves the device. The offline default is
tested in continuous integration by running the daemon inside a network namespace
with no connectivity.

\paragraph{Threat model.}
We summarize a STRIDE-style~\cite{shostack2014threat} analysis (Table~\ref{tab:threats}).
The trusted computing base is the local user account; the primary assets are the
in-flight transcript, the encrypted corpus and its key, and the local IPC endpoint.
The most consequential residual risks are \emph{same-uid} threats---a process
running as the same user can reach the IPC socket or the decrypted corpus---and
\emph{focus-hijack} injection, where a window that steals focus during dictation
receives the injected text. These are inherent to a user-level input tool and are
documented rather than hidden.

\begin{table}[t]
\centering
\small
\caption{Selected threats and \yazses{}' mitigations (STRIDE-style).}
\label{tab:threats}
\begin{tabular}{p{0.30\linewidth}p{0.60\linewidth}}
\toprule
Threat & Mitigation / residual risk \\
\midrule
Audio exfiltration & No network path by default; CI network-namespace gate. Residual: none on default path. \\
Corpus disclosure & AES-256-GCM protects the corpus \emph{at rest} (disk theft, backups). The key is machine-bound, so a same-uid process can derive it: encryption is not a same-uid boundary. \\
IPC abuse & Socket restricted to the owning user (0600). Residual: same-uid clients. \\
Wrong-window injection & Inject into the focused window only. Residual: focus hijack during a burst. \\
Model tampering & Models fetched from their canonical source on first run. Residual: supply-chain of the model host. \\
\bottomrule
\end{tabular}
\end{table}

\section{Privacy-Preserving On-Device Personalization}
\label{sec:personalization}

Dictation accuracy improves when the recognizer is biased toward a user's
own vocabulary and correction history, but the obvious way to personalize---send
corrections to a server---violates \yazses{}' invariants. \yazses{} instead keeps an
\emph{opt-in}, off-by-default learning loop entirely on-device. On each
hold-release, a background writer appends one event (transcript, optionally the
audio clip, and coarse metadata) to an encrypted SQLite corpus; text and audio are
encrypted with a machine-bound key while only coarse metadata stays in clear, and
capture never blocks the dictation hot path.

The \code{tune} command analyzes this corpus offline and \emph{proposes}
configuration changes---vocabulary additions, a VAD-threshold adjustment, a model
upgrade, disfluency tweaks---for the user to approve. It emits
configuration diffs, never model weights, and never transmits anything. To avoid
over-fitting proposals to the same data that generated them, the
tuner validates each proposal on a \emph{chronologically held-out} slice of the
corpus, with a guard against text that is duplicated across the split; proposals that
do not corroborate on held-out data are flagged as unverified. Correction signals
come from an explicit ``mark wrong'' action, from re-transcription with a larger
model as pseudo-ground-truth, and from a passive re-dictation heuristic---never from
keystroke logging.

This section contributes a \emph{design}: we describe the mechanism and its privacy
properties, not an accuracy evaluation. Quantifying how much the loop improves WER, and
confirming that the held-out guard rejects unhelpful proposals, requires longitudinal
single-user data and is future work (Section~\ref{sec:limitations}).

\section{Implementation}
\label{sec:impl}

\yazses{} is implemented in Python ($\geq$3.11) and packaged for PyPI, Snap, and an
APT repository. Heavy or platform-specific capabilities are isolated behind optional
install extras and imported lazily, so a base install stays small and every advanced
feature is dormant until explicitly enabled. The core is
\yzSrcFiles{} Python source files (\yzSLOC{} lines) with a test suite of
\yzTestFuncs{} test functions across \yzTestFiles{} files, and design decisions are
recorded in \yzADRs{} architecture decision records. The ``off-by-default, guarded''
extension pattern lets a large capability surface coexist with an unchanged default
dictation path.

\subsection{Capability surface}
\IfFileExists{figures/features_prose.tex}{\input{figures/features_prose.tex}}{%
  \newcommand{\yzFeatTotal}{\textbf{??}}\newcommand{\yzFeatDefault}{\textbf{??}}%
  \newcommand{\yzFeatRec}{\textbf{??}}\newcommand{\yzFeatOpt}{\textbf{??}}%
  \newcommand{\yzFeatExp}{\textbf{??}}\newcommand{\yzFeatDefaultNames}{\textbf{??}}%
  \newcommand{\yzFeatExpNames}{\textbf{??}}}
Beyond core dictation, \yazses{} registers \yzFeatTotal{} named capabilities in a
feature registry that drives the \code{yazses features} command; each carries a
recommendation tier, and Table~\ref{tab:features} summarizes the registry by
category. Only \yzFeatDefault{} capabilities are active on a fresh install
(\yzFeatDefaultNames{}); \yzFeatRec{} more are recommended, and the remaining
majority stay dormant until enabled. \yzFeatExp{} capabilities are marked
\emph{experimental}---the registry refuses to enable them without an explicit
\code{--force}, and none are part of the evaluation in this paper. The experimental
set comprises \yzFeatExpNames{}. Full per-feature documentation, including
configuration keys and dependencies, ships with the project.

\begin{table}[t]
\centering
\small
\caption{The capability registry by category and recommendation tier (generated from
the registry itself). ``Default'' capabilities are on after a fresh install;
``Experim[ental]'' ones refuse to enable without \code{--force} and are excluded from
this paper's evaluation.}
\label{tab:features}
\IfFileExists{figures/tab_features.tex}{\input{figures/tab_features.tex}}{(table
generated by \code{make\_features\_table.py})}
\end{table}

\section{Evaluation}
\label{sec:eval}

\paragraph{Setup.}
All measurements were taken on a single commodity laptop: \yzCPUModel{}
(\yzNumCPUs{} logical CPUs), \yzRAMGB{}~GB RAM, \yzOSName{}, Python~\yzPyVer{},
\code{faster-whisper}~\yzFWVer{}, CPU int8 (no GPU). WER is computed on
\yzWERn{} utterances spanning \yzWERspeakers{} speakers of
LibriSpeech~\cite{panayotov2015librispeech} \emph{test-clean}
(\yzWERaudioMin{}~minutes of audio), drawn by a deterministic
\emph{speaker-stratified} round-robin across all test-clean speakers so the subset is
reproducible without clustering on a few speakers. The standard Whisper English text
normalizer~\cite{whispernormalizer} is applied to both reference and hypothesis and
WER is computed by JiWER~\cite{jiwer}. Every number below is produced by the released
benchmark harness and stored as JSON with this provenance block.

\begin{figure}[t]
\centering
\includegraphics[width=0.9\linewidth]{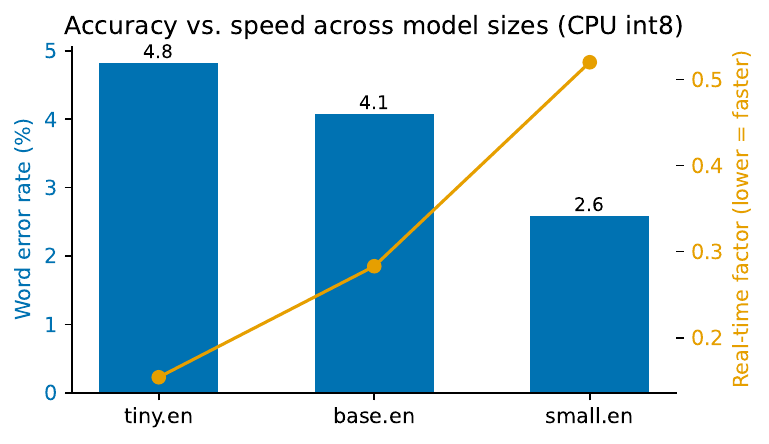}
\caption{Word error rate (bars) and median real-time factor (line) across model
sizes. Larger models are more accurate and slower, but all decode faster than real
time on CPU.}
\label{fig:wer}
\end{figure}

\paragraph{Accuracy and speed.}
Table~\ref{tab:main} and Figure~\ref{fig:wer} report WER and real-time factor (RTF,
decode time over audio duration; RTF${}<1$ is faster than real time) across the three
shipping model sizes. WER falls from \yzWERtiny\% (\code{tiny.en}) to \yzWERsmall\%
(\code{small.en}); all three models decode faster than real time on CPU, with
\code{small.en} at $\text{RTF}=\yzRTFsmall$ (about \yzSpeedSmall$\times$ real time)
and \code{tiny.en} at roughly \yzSpeedTiny$\times$. The table also reports cold-start
load time, resident memory, and on-disk model size.

Two caveats apply when reading these numbers. First, \yazses{} uses
Whisper \emph{unmodified}, so these figures characterize the shipping decode
configuration (CPU int8) rather than a new model; they check that the
on-device configuration reproduces the model's expected accuracy, and claim nothing
more.
For reference, the largest zero-shot Whisper model reports a 2.7\% word error rate on
LibriSpeech test-clean (after the same Whisper text normalization; Table~2
of~\cite{radford2023whisper}). On this clean read-speech benchmark the CPU-int8
\code{small.en} lands in the same range; the larger models' documented advantage is
robustness on harder, out-of-distribution audio~\cite{radford2023whisper}---the
conditions a real microphone actually produces. Second, this
measurement decodes clean recordings from file and therefore \emph{isolates the STT
stage}: it does not exercise the microphone, the VAD gate, pre-speech padding, or the
\code{initial\_prompt} biasing that live dictation adds (see Threats to validity).

\begin{table}[t]
\centering
\small
\caption{STT decode accuracy, speed, and footprint across model sizes (\yzWERn{}
LibriSpeech test-clean utterances; CPU int8; \yzCPUModel{}). RTF = decode time / audio
duration. These characterize the decode stage on clean read speech, not end-to-end
microphone dictation.}
\label{tab:main}
\input{figures/tab_main.tex}
\end{table}

\begin{figure}[t]
\centering
\includegraphics[width=0.9\linewidth]{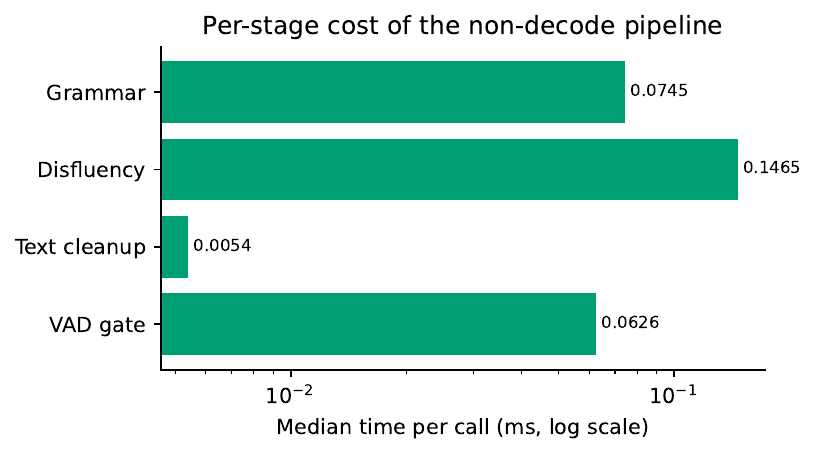}
\caption{Median per-call cost of the non-decode pipeline stages (log scale). Every
stage is well under a millisecond.}
\label{fig:latency}
\end{figure}

\paragraph{Pipeline overhead.}
The non-decode stages are negligible next to the decode. Table~\ref{tab:pipeline}
and Figure~\ref{fig:latency} give per-stage timings; the total non-decode overhead
per utterance is \yzOverhead~ms, i.e.\ the pipeline around the model adds far less
than a millisecond and end-to-end latency is dominated by the decode.

\begin{table}[t]
\centering
\small
\caption{Per-call cost of the pure (non-decode) pipeline stages.}
\label{tab:pipeline}
\input{figures/tab_pipeline.tex}
\end{table}

\paragraph{Command grammar.}
On \yzCmdN{} labelled command phrases the Tier-1 regex grammar achieves \yzCmdAcc\%
action accuracy, classifying \yzCmdAsCmd\% of them as commands (the remainder are
intentional dictation controls in the fixture). On \yzDictN{} lines of ordinary
prose, the false-positive rate---dictation wrongly treated as a command---is
\yzCmdFPR\%. Classification takes \yzCmdMs~ms per call, three orders of magnitude
below the perceptual budget, so the grammar is effectively free. These figures are
measured on text fixtures and therefore assume correct transcripts; real-world misfires
additionally depend on upstream ASR errors, so they upper-bound the grammar's own
contribution to command reliability rather than the end-to-end rate.

\paragraph{VAD gate.}
At the default threshold the RMS gate detects speech in \yzVADspeech\% of real
speech clips and rejects \yzVADsilence\% of silent/quiet-room buffers, with the
median speech level about \yzVADmargin$\times$ the threshold. The silence set is
synthetic sub-threshold noise, so the high rejection rate is unsurprising;
the informative result is the comfortable margin between real speech energy and the
threshold. This is a separability check at the operating point, not a frame-level
benchmark, and the RMS gate is not a substitute for a neural VAD in adverse noise (see
Section~\ref{sec:limitations}).

\paragraph{Dysfluency-friendly mode.}
\yazses{}' opt-in collapse pass for stuttered/repeated speech is held to a
pre-registered gate (thresholds fixed before measuring). On \yzDysNclean{} clean
control strings its false-collapse rate is \yzDysFP\%, and on \yzDysNdys{} labelled
dysfluency cases its recall is \yzDysRecall\%, meeting the pre-registered criteria
(false-collapse $<2\%$, recall $\geq 60\%$). This is measured on hand-authored text,
not affected-speaker audio, which remains future work.

\paragraph{Threats to validity.}
These numbers characterize one machine, one operating system, one English read-speech
corpus subset, and English \code{.en} models. All measurements were taken on Linux;
nothing here validates the macOS or Windows paths beyond their unit tests. WER on
read audiobooks is a floor, not a guarantee for noisy microphones or
accented/conversational speech; RTF scales with CPU; and the VAD result is a
separability check at the operating threshold, not a frame-level benchmark on a
diarized noisy corpus. We report the measurement conditions with every number so
results can be reproduced and re-scoped.

\section{Limitations and Future Work}
\label{sec:limitations}

\yazses{} ships English \code{.en} models by default; other languages require a
different model. It is a desktop tool with no mobile or web build. For the absolute
lowest WER on a difficult microphone, a cloud service may still win---the deliberate
trade-off is that nothing leaves the machine. Although the codebase targets three
operating systems, everything evaluated in this paper ran on Linux: the macOS and
Windows backends are implemented, covered by the unit suite in CI, and shipped as
unsigned developer-preview builds, but end-to-end dictation on those platforms has
not been validated by the author, and the cross-platform claim should be read as
architectural until it has been. The shipping system uses a single
push-to-talk decode; an exploratory branch investigated a streaming dual-stack STT
design and an on-device agentic layer (LLM tool-use with encrypted personal memory),
which we regard as future work rather than part of the evaluated system. Finally,
validating dysfluency-friendly mode and the personalization loop on real
affected-speaker and long-term single-user data is important future work.

The largest gap is that this paper reports \emph{no user study}. The evaluation is
entirely technical; usability, learnability, and the lived experience of the
accessibility features (dysfluency mode, EMG trigger, gaze targeting) are not
measured here. Claims about the system being easy to adopt therefore describe its
\emph{design intent}, not a measured outcome, and a human-subjects evaluation---in
particular with users who rely on the accessibility paths---is the most valuable next
step.

\section{Conclusion}

\yazses{} shows that voice dictation with practical voice commands can be delivered
entirely on-device---from a single codebase that targets Linux, macOS, and
Windows, and validated end-to-end on Linux---with privacy as an enforced invariant
rather than a promise. On commodity CPU hardware it transcribes faster than real
time at the word error rates of the underlying Whisper models, adds negligible
pipeline overhead, and---on text fixtures---classifies commands with no false
positives on ordinary prose, all while never sending audio or text off the machine.
We hope both the system and its reproducible benchmark harness are useful to
practitioners who need dictation they can trust.

\section*{Availability and Reproducibility}
\yazses{} is open-source under Apache-2.0 at
\url{https://github.com/MSKazemi/yazses} and on PyPI (\code{pipx install yazses}).
The benchmark harness that produced every number in Section~\ref{sec:eval}, together
with the machine-readable results, is released with the paper; each experiment
reproduces from a pinned dependency set and the public LibriSpeech
\emph{test-clean} download.

\bibliographystyle{plain}
\bibliography{refs}

\end{document}

%% file: figures/numbers.tex
\newcommand{\yzCPUModel}{13th Gen Intel(R) Core(TM) i7-1370P}
\newcommand{\yzNumCPUs}{20}
\newcommand{\yzRAMGB}{32}
\newcommand{\yzOSName}{Ubuntu 24.04.4 LTS}
\newcommand{\yzPyVer}{3.12.3}
\newcommand{\yzFWVer}{1.2.1}
\newcommand{\yzWERn}{200}
\newcommand{\yzWERspeakers}{40}
\newcommand{\yzWERaudioMin}{30.1}
\newcommand{\yzWERtiny}{4.82}

\newcommand{\yzWERsmall}{2.59}

\newcommand{\yzRTFsmall}{0.520}
\newcommand{\yzSpeedTiny}{6.5}
\newcommand{\yzSpeedSmall}{1.9}

\newcommand{\yzOverhead}{0.289}
\newcommand{\yzCmdAcc}{100}
\newcommand{\yzCmdAsCmd}{92.6}
\newcommand{\yzCmdFPR}{0.0}
\newcommand{\yzCmdMs}{0.021}
\newcommand{\yzCmdN}{68}
\newcommand{\yzDictN}{80}
\newcommand{\yzVADspeech}{100.0}
\newcommand{\yzVADsilence}{100.0}
\newcommand{\yzVADmargin}{2.4}
\newcommand{\yzDysFP}{0.0}
\newcommand{\yzDysRecall}{92.9}
\newcommand{\yzDysNclean}{33}
\newcommand{\yzDysNdys}{28}
\newcommand{\yzTestFiles}{201}
\newcommand{\yzTestFuncs}{1,639}
\newcommand{\yzADRs}{147}
\newcommand{\yzSLOC}{30,306}
\newcommand{\yzSrcFiles}{403}

%% file: figures/features_prose.tex
\newcommand{\yzFeatTotal}{136}
\newcommand{\yzFeatDefault}{4}
\newcommand{\yzFeatRec}{15}
\newcommand{\yzFeatOpt}{105}
\newcommand{\yzFeatExp}{12}
\newcommand{\yzFeatDefaultNames}{Dictation core, Voice commands, Mid-Thought Undo, and Voice-activity overlay}
\newcommand{\yzFeatExpNames}{Wake-Word Activation, HatSelect Structural Editing, Vocal Morse, Mouth-Sound Switch Access, Pitch-Contour Gestures, Vocal Joystick, Modality Role Router, Glance-Type (camera), Crowd-Proof Dictation, Cocktail Filter (voice focus), Voice Guard (biometric + anti-spoof), and Glasses--Desktop Bridge}

%% file: figures/tab_features.tex
\begin{tabular}{lrrrrr}
\toprule
Category & Total & Default & Recomm. & Opt-in & Experim. \\
\midrule
Core dictation & 16 & 4 & 3 & 8 & 1 \\
Accuracy \& correction & 20 & 0 & 6 & 14 & 0 \\
Formatting \& structure & 31 & 0 & 3 & 28 & 0 \\
Editing \& navigation & 13 & 0 & 1 & 11 & 1 \\
Commands \& automation & 12 & 0 & 2 & 10 & 0 \\
Multilingual & 7 & 0 & 0 & 7 & 0 \\
Accessibility \& input modalities & 20 & 0 & 0 & 14 & 6 \\
Learning, memory \& analytics & 9 & 0 & 0 & 9 & 0 \\
Conversation \& recording capture & 8 & 0 & 0 & 4 & 4 \\
\midrule
\textbf{All categories} & \textbf{136} & 4 & 15 & 105 & 12 \\
\bottomrule
\end{tabular}

%% file: figures/tab_main.tex
\begin{tabular}{lrrrrrr}
\toprule
Model & WER & RTF & Speed & Load & RSS & Disk \\
 & (\%) & (med.) & & (s) & (MB) & (MB) \\
\midrule
tiny.en & 4.82 & 0.154 & 6.5$\times$ & 0.6 & 37 & 78 \\
base.en & 4.07 & 0.283 & 3.5$\times$ & 0.8 & 53 & 148 \\
small.en & 2.59 & 0.520 & 1.9$\times$ & 1.6 & 520 & 486 \\
\bottomrule
\end{tabular}

%% file: figures/tab_pipeline.tex
\begin{tabular}{lrr}
\toprule
Pipeline stage & Median (ms) & P95 (ms) \\
\midrule
VAD gate (RMS) & 0.0626 & 0.0756 \\
Text cleanup & 0.0054 & 0.0057 \\
Disfluency filter & 0.1465 & 0.1701 \\
Grammar classify (dictation) & 0.0745 & 0.0931 \\
Grammar classify (command) & 0.0435 & 0.0489 \\
\midrule
\textbf{Total non-decode overhead} & \textbf{0.289} & \\
\bottomrule
\end{tabular}

%% file: main.bbl
\begin{thebibliography}{10}

\bibitem{whispernormalizer}
Kurian Benoy.
\newblock whisper-normalizer: A python package for text standardization.
\newblock \url{https://github.com/kurianbenoy/whisper_normalizer}, 2024.
\newblock Implements the {Whisper} {EnglishTextNormalizer}. Accessed July 2026.

\bibitem{desplanques2020ecapa}
Brecht Desplanques, Jenthe Thienpondt, and Kris Demuynck.
\newblock {ECAPA-TDNN}: Emphasized channel attention, propagation and
  aggregation in {TDNN} based speaker verification.
\newblock In {\em Proc. Interspeech 2020}, pages 3830--3834, 2020.

\bibitem{llamacpp}
Georgi Gerganov et~al.
\newblock llama.cpp: {LLM} inference in {C/C++}.
\newblock \url{https://github.com/ggml-org/llama.cpp}, 2024.
\newblock GBNF grammar-constrained decoding. Accessed July 2026.

\bibitem{jeffries2024moonshine}
Nat Jeffries, Evan King, Manjunath Kudlur, Guy Nicholson, James Wang, and Pete
  Warden.
\newblock Moonshine: Speech recognition for live transcription and voice
  commands, 2024.

\bibitem{windowsvoiceaccess}
{Microsoft}.
\newblock Use voice access to control your {PC} and author text with your
  voice.
\newblock
  \url{https://support.microsoft.com/en-us/topic/use-voice-access-to-control-your-pc-author-text-with-your-voice-4dcd23ee-f1b9-4fd1-bacc-862ab611f55d},
  2024.
\newblock Accessed July 2026.

\bibitem{msrecall}
{Microsoft}.
\newblock Retrace your steps with {Recall}.
\newblock
  \url{https://support.microsoft.com/en-us/windows/retrace-your-steps-with-recall-aa03f8a0-a78b-4b3e-b0a1-2eb8ac48701c},
  2025.
\newblock Feature announced May 2024; general availability December 2025.
  Accessed July 2026.

\bibitem{dragon}
{Nuance Communications (Microsoft)}.
\newblock Dragon speech recognition software and solutions.
\newblock \url{https://dragon.nuance.com/}, 2026.
\newblock Dragon Professional for Mac discontinued October 2018. Accessed July
  2026.

\bibitem{ctranslate2}
{OpenNMT}.
\newblock {CTranslate2}: Fast inference engine for transformer models.
\newblock \url{https://github.com/OpenNMT/CTranslate2}, 2024.
\newblock Accessed July 2026.

\bibitem{panayotov2015librispeech}
Vassil Panayotov, Guoguo Chen, Daniel Povey, and Sanjeev Khudanpur.
\newblock Librispeech: An {ASR} corpus based on public domain audio books.
\newblock In {\em 2015 IEEE International Conference on Acoustics, Speech and
  Signal Processing (ICASSP)}, pages 5206--5210. IEEE, 2015.

\bibitem{radford2023whisper}
Alec Radford, Jong~Wook Kim, Tao Xu, Greg Brockman, Christine McLeavey, and
  Ilya Sutskever.
\newblock Robust speech recognition via large-scale weak supervision.
\newblock In {\em Proceedings of the 40th International Conference on Machine
  Learning (ICML)}, pages 28492--28518, 2023.
\newblock arXiv:2212.04356.

\bibitem{ydotool}
{ReimuNotMoe}.
\newblock ydotool: Generic {Linux} command-line automation tool.
\newblock \url{https://github.com/ReimuNotMoe/ydotool}, 2024.
\newblock Accessed July 2026.

\bibitem{shostack2014threat}
Adam Shostack.
\newblock {\em Threat Modeling: Designing for Security}.
\newblock Wiley, 2014.

\bibitem{sileovad}
{Silero Team}.
\newblock Silero {VAD}: pre-trained enterprise-grade voice activity detector.
\newblock \url{https://github.com/snakers4/silero-vad}, 2024.
\newblock MIT license. Accessed July 2026.

\bibitem{fasterwhisper}
{SYSTRAN}.
\newblock faster-whisper: Faster {Whisper} transcription with {CTranslate2}.
\newblock \url{https://github.com/SYSTRAN/faster-whisper}, 2024.
\newblock Accessed July 2026.

\bibitem{talon}
{Talon Voice}.
\newblock Talon voice.
\newblock \url{https://talonvoice.com}, 2024.
\newblock Accessed July 2026.

\bibitem{jiwer}
Nik Vaessen.
\newblock {JiWER}: Similarity measures for automatic speech recognition
  evaluation.
\newblock \url{https://github.com/jitsi/jiwer}, 2024.
\newblock Accessed July 2026.

\bibitem{wisprflow}
{Wispr AI}.
\newblock Wispr flow.
\newblock \url{https://wisprflow.ai}, 2024.
\newblock Accessed July 2026.

\end{thebibliography}
